\documentclass[hyper]{JHEP} 

\usepackage{epsfig}

\newcommand\fverb{\setbox\pippobox=\hbox\bgroup\verb}

\newcommand\fverbdo{\egroup\medskip\noindent%

            \fbox{\unhbox\pippobox}\ }

\newcommand\fverbit{\egroup\item[\fbox{\unhbox\pippobox}]}

\newbox\pippobox

\title{Fundamental
 String  and D1-brane   in I-brane Background}
\author{by J. Kluso\v{n}\\
     Department of Theoretical Physics and Astrophysics\\
                   Faculty of Science, Masaryk University\\
Kotl\'{a}\v{r}sk\'{a} 2, 611 37, Brno\\
Czech Republic\\
    E-mail: \email{klu@physics.muni.cz}}
\preprint{arXiv:0711.4219v2}

 \abstract{This paper is devoted to the study of dynamics of fundamental
string and D1-brane in I-brane
background. We consider configurations
where string and D1-brane uniformly
wrap transverse spheres. We explicitly
determine corresponding conserved
charges and find relations between
them.}


\def\mF{\mathcal{F}}

\def\bM{\mathbf{M}}

\def\bz{\mathbf{z}}
\def\by{\mathbf{y}}

\def\bA{\mathbf{A}}

\def\bAi{\left(\mathbf{A}^{-1}\right)}

\def\mL{\mathcal{L}}

\def\mJ{\mathcal{J}}

\begin{document}
\section{Introduction and Summary}\label{first}
Study of different configurations of
branes provides us information about
their dynamics and  properties of gauge
theories that reside in them. One such
an interesting configuration of space
filling intersecting D-branes are
D5-branes intersecting over $1+1$
dimensions \cite{Green:1996dd}. The
characteristic property of such a
configurations is that they describe
chiral theories in intersection domain.
Further, as was observed in
\cite{Itzhaki:2005tu}
 there is an
interesting symmetry   enhancement
where the Poincare symmetry is
enhancements from $SO(1,1)$ to
$SO(1,2)$ and the number of
supersymmetries is doubled
\footnote{For some relevant works, see
\cite{Hung:2006nn,Berg:2006ng,Hung:2006jh,Grisa:2006tm,Antonyan:2006pg,
Antonyan:2006qy,Antonyan:2006vw,Kluson:2006wa,
Kluson:2005qq,Kluson:2005eb} .}. In
fact, using the standard ideas of
holography one could expect that  the
dynamics at the $1+1$ dimensional
intersection of the two sets of
fivebranes should be holographically
related to a $2 + 1$ dimensional bulk
theory, with the extra dimension being
the radial direction away from the
intersection. However as was nicely
shown in \cite{Itzhaki:2005tu}  the
bulk description includes two radial
directions away from each set of
fivebranes, and is $3 + 1$ dimensional
and hence the corresponding boundary
theory is $2 + 1$ dimensional. More
precisely, the strong coupling limit of
the system is described as a stack of
intersecting NS5-branes.  When all the
fivebranes are coincident, the
near-horizon geometry is
\begin{equation}\label{exactsol}
R^{2,1}\times
 R_{\phi} \times  SU(2)_{k_1} \times
 SU(2)_{k_2} \ .
\end{equation}
 Here, $R_{\phi}$ is one combination of the
radial directions away from the two
sets of fivebranes, and the coordinates
of $R^{2,1}$  are $x^0,x^1$ and another
combination of the two radial
directions. The two $SU(2)$'s describe
the angular three-spheres corresponding
to $(R^4)_{2345}$  and $(R^4)_{6789}$.
The fact that (\ref{exactsol}) is an
exact solution of the classical string
theory equations of motion allows us to
obtain information about the
intersecting fivebrane system, which is
not accessible via a gauge theory
analysis.

As it is clear from (\ref{exactsol})
the geometry (\ref{exactsol}) exhibits
a higher symmetry than the full brane
configuration. In particular, the
combination of radial directions away
from the intersection that enters
$R^{2,1}$ appears symmetrically with
the other spatial direction, and the
background has a higher Poincare
symmetry, $ISO(2, 1)$, than the
expected $ISO(1, 1)$ \footnote{An
important issue is whether  this higher
symmetry is an exact property of string
theory in the background
(\ref{exactsol}), or whether it is
broken by quantum effects. This issue
was carefully discussed in
\cite{Itzhaki:2005tu}.}.

This background geometry has many
unusual properties and certainly
deserves better understanding. The
holographic mapping between field
theory living on the I-brane
intersection and the bulk theory were
studied in \cite{Itzhaki:2005tu}. On
the other hand it is also well  known
from the study of $AdS/CFT$
correspondence that it is possible to
derive
 information about the boundary  CFT theory
from the study of the semiclassical
string and D1-brane configurations in
the bulk of $AdS_5$
 \footnote{For review, see
\cite{Tseytlin:2003ii,Tseytlin:2004xa,Plefka:2005bk}.}.

These recent developments give us the
motivation to investigate semiclassical
strings in I-brane background as well.
In this paper we would like to
investigate multispin rotating string
and D1-brane solutions, having angular
momenta in both three-spheres  of
I-brane background. Our goal is to find
relation between energy and
corresponding conserved charges.
Unfortunately due to the limited amount
of information about dual world-volume
I-brane theory we will not be able to
compare them with their dual
counterparts. On the other hand we mean
that it is useful exercise to study the
dynamics of fundamental string and
D1-brane in such a non-trivial
background.

In fact, it  turns out that  the
presence of a non-trivial NS two form
field has an important  consequences
for the dynamics of string or D-brane.
Explicitly, in order to find
 relation between
energy and conserved charges we have to
consider  situation when string moves
on  both  spheres $S^{3}$
simultaneously. Then we find that the
dependence of energy on the conserved
charges takes  similar form as an
ordinary relativistic relation between
energy and momenta with the exception
that the conserved charges related to
the motion on the second three-sphere
are functions of the charges  related
to the motion on the first three
sphere.

As the next step we consider the motion
of probe D1-brane in I-brane
background. Study of homogeneous
configurations of probe Dp-brane in
similar  background has been performed
in the past  in many papers
\cite{Kutasov:2005rr,Thomas:2005fw,Kluson:2005jr,Thomas:2005am,Chen:2005wm,
Kluson:2005qx,Thomas:2004cd,Bak:2004tp,Kluson:2004yk,Kluson:2004xc,
Kutasov:2004ct,Sahakyan:2004cq,
Ghodsi:2004wn,Burgess:2003mm,
Panigrahi:2004qr,Nakayama:2004yx}
\footnote{For review and extensive list
of references, see
\cite{Nakayama:2007sb}.} .
 Our approach can be considered as
generalization of this approach when we
study D1-brane with spatial dependent
world-sheet modes. We  solve equations
of motion and we find explicit time
dependence of the radial mode in given
background. Surprisingly if we impose
the condition that D1-brane  radial
velocity is zero we derive  conditions
that have the same form as Virasoro
conditions in case of the fundamental
string. Then we will be able to find
relation between energy and conserved
momenta that again has similar
structure as the  relation  that was
derived in case of fundamental string.
It would be certainly interesting to
study dual theory living on I-brane and
try to find corresponding states that
are dual to the classical fundamental
string and D1-brane configurations.

As the possible extension of our work
we suggest to study  giant magnon and
spike configurations on fundamental
string in I-brane background following
recent works
\cite{Hofman:2006xt,Ishizeki:2007we,Bobev:2007bm,
Chu:2006ae}. We hope to return to this
problem in future.

 The organization of this
work is as follows. In the next section
(\ref{second}) we review properties of
I-brane background. In section
(\ref{third}) we study the classical
string in given background and we find
solutions of the equation of motions.
In section (\ref{fourth}) we study
D1-brane probe in given background and
we find exact solutions of the equation
of motion with some interesting
properties.
\section{Review of  I-brane Background}\label{second}
In this section we review
 the  background studied in
 work \cite{Itzhaki:2005tu}. This
background is known as I-brane
background  and arises from the
configurations of intersecting
NS5-branes. Namely, we consider  the
intersection of two stack of
NS5-branes. We have $k_1$ NS5-branes
 extended in $(0,1,2,3,4,5)$
directions and the set of $k_2$
NS5-branes extended in $(0,1,6,7,8,9)$
directions. Let us define
\begin{eqnarray}
\by&=&(x^2,x^3,x^4,x^5) \ , \nonumber \\
\bz&=&(x^6,x^7,x^8,x^9) \
\nonumber \\
\end{eqnarray}
and presume that we have $k_1$
NS5-branes localized at the points
$\bz=0$ and $k_2$ NS5-branes localized
at the points $\by=0$. The supergravity
background corresponding to this
configuration takes the form
\begin{eqnarray}\label{bg}
\Phi(\bz,\by)&=&\Phi_1(\bz)+
\Phi_2(\by) \ , \nonumber \\
g_{\mu\nu}&=&\eta_{\mu\nu} \ , \quad
\mu,\nu=0,1 \ , \nonumber \\
g_{\alpha\beta}&=&e^{2(\Phi_2- \Phi_2(\infty))}\delta_{\alpha\beta}
\ , \quad  \mathcal{H}_{\alpha\beta\gamma}=
-\epsilon_{\alpha\beta\gamma\delta}
\partial^\delta \Phi_2 \ , \quad
\alpha,\beta,\gamma,\delta=
2,3,4,5 \ , \nonumber \\
g_{pq}&=&e^{2(\Phi_1-\Phi_1(\infty))} \delta_{pq} \ , \quad
\mathcal{H}_{pqr}=-\epsilon_{pqrs}
\partial^s\Phi_1 \ , \quad
p,q,r,s=6,7,8,9 \ , \nonumber \\
\end{eqnarray}
where $\Phi$ on the
first line means the dilaton
and where
\begin{eqnarray}
e^{2(\Phi_1-\Phi_1(\infty))}=H_1(\bz)\equiv
\frac{\lambda_1} {r^2_1} \ , \quad
 \lambda_1=k_1l_s^2 \ , \nonumber \\
 e^{2(\Phi_2-\Phi_2(\infty))}=H_2(\by)
 \equiv
\frac{\lambda_2} {r^2_2} \ , \quad
\lambda_2=k_2l_s^2 \ , \nonumber \\
\end{eqnarray}
where we consider the near horizon
limit of given background.
Then the metric takes the form
\begin{eqnarray}\label{NS5bac}
ds^2=-dt^2+dx_1^2+\frac{\lambda_1}{r_1^2}dr_1^2+
\frac{\lambda_2}{r_2^2}dr_2^2+
\lambda_1d\Omega^{(3)}_1+
\lambda_2d\Omega^{(3)}_2  \ ,
\end{eqnarray}
where $d\Omega_1^{(3)}$ and
$d\Omega_2^{(3)}$ correspond to the
line elements on the unit three sphere
in the form
\begin{eqnarray}
d\Omega^{(3)}_1&=&
 d\theta^2_1+ \sin^2\theta_1 d \phi_1^2+
 \cos^2\theta_1 d\psi_1^2 \  , \nonumber \\
  0 &<& \theta_1 < \frac{\pi}{2} \ , \quad  0 = \phi_1 \ , \quad \psi_1 < 2\pi
 \ ,   \nonumber \\
d\Omega^{(3)}_2&=& d\theta^2_2+ \sin^2\theta_2 d \phi_2^2+
 \cos^2\theta_2 d\psi_2^2 \  , \nonumber \\
  0 &<& \theta_2 < \frac{\pi}{2}\ ,  \quad 0 = \phi_2\  , \quad
  \psi_2 < 2\pi \ .
  \nonumber \\
\end{eqnarray}
Finally, we choose the gauge where
the non-zero components of
 NS two form take the form
\begin{eqnarray}
 b^{(1)}_{\phi_1\psi_1}=
   \lambda_1\cos^2\theta_1 \ ,\quad
 b^{(2)}_{\phi_2\psi_2}=
   \lambda_2\cos^2\theta_2 \ .
 \nonumber \\
\end{eqnarray}
\section{Fundamental String in I-brane
Background}\label{third} In this
section we study dynamics of
fundamental string in I-brane
background.
 Our starting point is the Polyakov form of the string
action in general  background
\begin{eqnarray}\label{actPol}
S&=&-\frac{1}{4\pi \alpha'}
\int_{-\pi}^\pi d\sigma d\tau
[\sqrt{-\gamma}\gamma^{\alpha\beta}
g_{MN}\partial_\alpha x^M\partial_\beta x^N
-e^{\alpha\beta}
\partial_\alpha x^M\partial_\beta x^N b_{MN}]+
\nonumber \\
&+&\frac{1}{4\pi}\int_{-\pi}^\pi
d\sigma d\tau \sqrt{-\gamma}
R\Phi \ , \nonumber \\
\end{eqnarray}
 where $\gamma^{\alpha\beta}$ is
a world-sheet metric, $R$ is
 its Ricci scalar and
 $e^{\tau\sigma}=-\epsilon^{\sigma\tau}=1$.
 Finally,  the modes
$x^M \ , M=0,\dots,9$ parameterize the
embedding of the string in given
background.
 The variation of the action (\ref{actPol})
with respect to $x^K$ implies following
equations of motion
\begin{eqnarray}\label{eqgx}
& &-\frac{1}
{4\pi\alpha'}\sqrt{-\gamma}\gamma^{\alpha\beta}
\partial_K
g_{MN}\partial_\alpha x^M\partial_\beta x^N
+\frac{1}{2\pi\alpha'}\partial_\alpha[
\sqrt{-\gamma}\gamma^{\alpha\beta}
g_{KM}\partial_\beta x^M]-
\nonumber \\
&-&\frac{1}{2\pi\alpha'}
\partial_\alpha[
\epsilon^{\alpha\beta}
\partial_\beta x^M b_{KM}]
+\frac{1}{4\pi\alpha'}
\epsilon^{\alpha\beta}
\partial_\alpha x^M\partial_\beta x^N
\partial_K b_{MN}
+\frac{1}{4\pi}\partial_K
\Phi \sqrt{-\gamma}R=0 \ .
\nonumber \\
\end{eqnarray}
Further, the variation of the action with
respect to the metric components
 imply the constraints
\begin{eqnarray}\label{gravcons}
& &
T_{\alpha\beta}\equiv-\frac{4\pi}{\sqrt{-\gamma}}
\frac{\delta S}{\delta \gamma^{\alpha
\beta}}= \frac{1}{\alpha'}
g_{MN}\partial_\alpha x^M\partial_\beta
x^N-R_{\alpha \beta}+
\nonumber \\
&+&(\nabla_\alpha \nabla_\beta x^M)
\partial_M \Phi+(\partial_\alpha x^M\partial_\beta
x^N)\partial_M\partial_N\Phi
-\nonumber \\
&-&\frac{1}{2}
\gamma_{\alpha\beta}
\left(\frac{1}{\alpha'}
\gamma^{\gamma\delta}
\partial_\gamma x^M\partial_\delta
x^N g_{MN}-R\Phi+2
\nabla^\alpha \nabla_\alpha \Phi\right) \ .
\nonumber \\
\end{eqnarray}
As the first step we  introduce two
modes $\rho_1$ and $\rho_2$ defined as
\begin{equation}\label{RSp}
r_1=e^{\frac{\rho_1}{\sqrt{\lambda_1}} }\
, \quad
r_2=e^{\frac{\rho_2}{\sqrt{\lambda_2}}} \ .
\end{equation}
In  the second step,   following
\cite{Itzhaki:2005tu} we
 introduce two modes $r,
y$ through the relation
\begin{equation}\label{phix2}
Qr=\frac{1}{\sqrt{\lambda_1}}\rho_1+
\frac{1}{\sqrt{\lambda_2}}\rho_2 \ , \quad
Qy=\frac{1}{\sqrt{\lambda_2}}\rho_1-
\frac{1}{\sqrt{\lambda_1}}\rho_2 \ ,
\end{equation}
where
\begin{equation}
Q=\frac{1}{\sqrt{\lambda}} \ , \quad
\frac{1}{\lambda}=\frac{1}{\lambda_1}+
\frac{1}{\lambda_2} \ .
\end{equation}
Using these transformations it is easy
to see that the dilaton $\Phi$
depends on $r$ only
\begin{eqnarray}
\Phi=\Phi_1+\Phi_2=
-Qr+\Phi_0 \ ,
\nonumber \\
\end{eqnarray}
where $\Phi_0\equiv \Phi_1(\infty)+
\Phi_2(\infty)$.
With the help of the variables $r,y$ the action
for string in $I$-brane background takes the form
\begin{eqnarray}\label{actPol2}
S&=&-\frac{1}{4\pi \alpha'}
\int_{-\pi}^\pi d\sigma d\tau
[\sqrt{-\gamma}\gamma^{\alpha\beta}
(-\partial_\alpha t\partial_\beta t
+\partial_\alpha r\partial_\beta r+
\partial_\alpha y\partial_\beta y+\nonumber \\
&+&g_{mn}\partial_\alpha x^m\partial_\beta x^n)
-e^{\alpha\beta}
\partial_\alpha x^m\partial_\beta x^n b_{mn}]-
\nonumber \\
&-&\frac{1}{4\pi}\int_{-\pi}^\pi
d\sigma d\tau \sqrt{-\gamma}
RQr \ , \nonumber \\
\end{eqnarray}
where $x^{m}$ label angular coordinates
corresponding to $S^3_1,S^3_2$
respectively. In the   conformal gauge
$
\gamma_{\alpha\beta}=\eta_{\alpha\beta}$
the constraints (\ref{gravcons}) that
follow from the variation of the action
(\ref{actPol2}) take
  simpler form
\begin{eqnarray}\label{Tcon}
T_{\sigma\sigma}&=&
\frac{1}{2\alpha'}
(-\partial_\sigma t\partial_\sigma t
-\partial_\tau t\partial_\tau t
+\partial_\sigma r\partial_\sigma r
+\partial_\tau r\partial_\tau r
+\partial_\sigma y\partial_\sigma y
+\partial_\tau y\partial_\tau y+
\nonumber \\
&+&g_{mn}\partial_\sigma x^m\partial_\sigma x^n
+g_{mn}\partial_\tau x^m\partial_\tau x^n)-
Q\partial_\tau^2r \ ,
\nonumber \\
T_{\tau\tau}&=&
\frac{1}{2\alpha'}
(-\partial_\sigma t\partial_\sigma t
-\partial_\tau t\partial_\tau t
+\partial_\sigma r\partial_\sigma r
+\partial_\tau r\partial_\tau r
+\partial_\sigma y\partial_\sigma y
+\partial_\tau y\partial_\tau y+
\nonumber \\
&+&g_{mn}\partial_\sigma x^m\partial_\sigma x^n
+g_{mn}\partial_\tau x^m\partial_\tau x^n)-
Q\partial_\sigma^2 r \ ,
\nonumber \\
T_{\tau\sigma}&=&
\frac{1}{\alpha'}(
-\partial_\tau t\partial_\sigma t
+\partial_\tau r\partial_\sigma r
+\partial_\tau y\partial_\sigma y
+g_{mn}\partial_\sigma x^m\partial_\tau x^n)
-Q\partial_\sigma\partial_\tau r \ .
\nonumber \\
\end{eqnarray}
Looking on the form of the background
(\ref{NS5bac})
we observe that the action (\ref{actPol2})
 is invariant under
following transformations of fields
\begin{eqnarray}
t'(\tau,\sigma)&=&t(\sigma,\tau)+\epsilon_t \ ,
\nonumber \\
y'(\tau,\sigma)&=&y(\tau,\sigma)+\epsilon_y  \ ,
\nonumber \\
\psi'_{1}(\tau,\sigma)&=&
\psi_{1}(\tau,\sigma)+\epsilon_{\psi_1} \ ,
\nonumber \\
\psi'_{2}(\tau,\sigma)&=&
\psi_{2}(\tau,\sigma)+\epsilon_{\psi_2} \ ,
\nonumber \\
\phi'_{1}(\tau,\sigma)&=&
\phi_{1}(\tau,\sigma)+\epsilon_{\phi_1} \ ,
\nonumber \\
\phi'_{2}(\tau,\sigma)&=&
\phi_{2}(\tau,\sigma)+\epsilon_{\phi_2} \ ,
\nonumber \\
\end{eqnarray}
where $\epsilon_t,\epsilon_y,\epsilon_{\phi_{1,2}},
\epsilon_{\psi_{1,2}}$
 are constants.
Then it is simple task to determine corresponding
conserved charges
\begin{eqnarray}\label{CGcon}
P_t&=&\frac{1}{2\pi\alpha'}
\int_{-\pi}^\pi d\sigma
\partial_\tau t \ , \quad
P_{y}=- \frac{1}{2\pi\alpha'}
\int_{-\pi}^\pi d\sigma
\partial_\alpha y \ ,
\nonumber \\
P_{\psi_1}&=& -\frac{1}{2\pi\alpha'}
\int_{-\pi}^\pi d\sigma [
g_{\psi_1\psi_1}\partial_\tau \psi_1 +
b_{\psi_1\phi_1}\partial_\sigma\phi_1]
\ ,
\nonumber \\
P_{\psi_2}&=& -\frac{1}{2\pi\alpha'}
\int_{-\pi}^\pi d\sigma [
g_{\psi_2\psi_2}\partial_\tau \psi_2 +
b_{\psi_2\phi_2}\partial_\sigma\phi_2]
\ ,
\nonumber \\
P_{\phi_1}&=& -\frac{1}{2\pi\alpha'}
\int_{-\pi}^\pi d\sigma [
g_{\phi_1\phi_1}\partial_\tau \phi_1 +
b_{\phi_1\psi_1}\partial_\sigma\psi_1]
\ ,
\nonumber \\
P_{\phi_2}&=& -\frac{1}{2\pi\alpha'}
\int_{-\pi}^\pi d\sigma [
g_{\phi_2\phi_2}\partial_\tau \phi_2 +
b_{\phi_2\psi_2}\partial_\sigma\psi_2]
\ .
\nonumber \\
\end{eqnarray}
Note that $P_t$ is related to the energy as
$P_t=-E$.

Our goal is to study the dynamics of
string that moves and wraps two spheres
$S^{3}_{1,2}$ simultaneously.
Explicitly, let us consider following
ansatz
\begin{eqnarray}\label{ansatz}
r&=&r(\tau), \quad  y=y(\tau), \quad
t=\kappa \tau ,
\nonumber \\
\theta_1&=&\theta_1^c\equiv \mathrm{const} \ , \quad
\psi_1=\omega_1\tau+n_1\sigma \ , \quad  \phi_1=\nu_1\tau+m_1\sigma  \ ,
\nonumber \\
\theta_2&=&\theta_2^c\equiv
\mathrm{const}, \quad \psi_2=
\omega_2\tau+n_2\sigma \ , \quad
\phi_2= \nu_2\tau+ m_2\sigma \ .
\nonumber \\
\end{eqnarray}

Let us now study the equations of
motion (\ref{eqgx}) for string moving
in the background (\ref{NS5bac}).
  In conformal gauge the
equations of motion for $t,y,r$ take the form
\begin{eqnarray}
\partial_\alpha[\eta^{\alpha\beta}\partial_\beta
t]=0 \ , \quad
\partial_\alpha[\eta^{\alpha\beta}\partial_\beta
r]=0 \ , \quad
\partial_\alpha[\eta^{\alpha\beta}\partial_\beta
y]=0 \
\end{eqnarray}
that we solve as
\begin{equation}
t=\kappa \tau \ , \quad
 r=v_r \tau+r_0
\ , \quad y=v_y\tau+y_0 \ .
\end{equation}
Further, the equations of motion for
 $\theta_{1}$ and $\theta_{2}$ for constant $\theta$'s
take the form
\begin{eqnarray}\label{eqtheta1}
\frac{\lambda_1\sin\theta^c_1\cos\theta^c_1}{2\pi\alpha'}
[ \nu_1^2-m_1^2-\omega_1^2+n_1^2
+2\omega_1 m_1-2\nu_1 n_1] =0 \ ,
\nonumber \\
\end{eqnarray}
\begin{eqnarray}\label{eqtheta2}
\frac{\lambda_2\sin\theta_2^c\cos\theta_2^c}
{2\pi\alpha'} [
\nu_2^2-m_2^2-\omega_2^2+n_2^2
+2\omega_2 m_2-2\nu_2 n_2] =0 \ .
\nonumber \\
\end{eqnarray}

 Finally,
the equation of motion for $\phi_1$ in
I-brane  background takes the form
\begin{eqnarray}
\frac{1}{2\pi\alpha'}
\partial_\alpha[\eta^{\alpha\beta}
g_{\phi_1\phi_1}\partial_\beta \phi_1]
-\frac{1}{2\pi\alpha'}
\partial_\alpha[\epsilon^{\alpha\beta}
\partial_\beta\psi_1 b_{\phi_1\psi_1}]=0 \
\nonumber \\
\end{eqnarray}
that is automatically obeyed for
(\ref{ansatz}). In the same way we can
show that the equations of motion for
$\psi_1,\phi_2,\psi_2$ are obeyed with the
ansatz (\ref{ansatz}). Further,
it is easy to see that for
\begin{eqnarray}\label{omega12}
\omega_1&=&\nu_1 \ , \quad m_1=n_1 \ , \nonumber \\
\omega_2&=&\nu_2 \ , \quad m_2=n_2 \  \nonumber \\
\end{eqnarray}
  the equations of
motion (\ref{eqtheta1}), (\ref{eqtheta2}) are satisfied
as well.

As the next step  we are going to analyze  Virasoro
constraints (\ref{Tcon}). The
constraint $T_{\tau\sigma}=0$ implies
\begin{eqnarray}\label{Tts}
\lambda_1\sin^2\theta^c_1 \nu_1 m_1+\lambda_1\cos^2\theta^c_1
\omega_1 n_1+\lambda_2\sin^2\theta^c_2\nu_2
m_2+\lambda_2\cos^2\theta^c_2
\omega_2 n_2=0 \ . \nonumber \\
\end{eqnarray}
If we insert (\ref{omega12}) into
(\ref{Tts}) we obtain the condition
\begin{equation}
\frac{\lambda_1}{\lambda_2}=-\frac{\omega_2 m_2}{\omega_1 m_1}
\end{equation}
that can be solved as
\begin{equation}
m_1=-m_2 \ , \quad \omega_2=\frac{\lambda_1}{\lambda_2}\omega_1 \
.
\end{equation}
Note that this solution is valid for any constant $\theta_{1,2}^c$.
 On the other hand the Virasoro constraints
$T_{\tau\tau}=T_{\sigma\sigma}=0$ imply
\begin{eqnarray}\label{Tss}
-\kappa^2+v_r^2+v_y^2+
\lambda_1(\omega_1^2+m_1^2)+
\lambda_2(\omega_2^2+m_2^2)=0 \ . \nonumber \\
\end{eqnarray}
 Let us now determine the
form of the conserved charges
$P_t,P_y,P_{\phi_1},P_{\phi_2},P_{\psi_1}$
and $P_{\psi_2}$ for the ansatz
(\ref{ansatz})
\begin{eqnarray}
P_t&=&\frac{1}{\alpha'}\kappa \ , \quad
P_y=-\frac{1}{\alpha'}v_y \ ,  \nonumber \\
 P_{\psi_1}&=&\frac{\lambda_1 }{\alpha'}
 [-\cos^2\theta_1^c\omega_1+\cos^2\theta_1^c m_1 ] \ , \nonumber \\
 P_{\psi_2}&=&\frac{\lambda_2 }{\alpha'}
 [-\cos^2\theta_2^c\omega_2+\cos^2\theta_2^c m_2 ] \ , \nonumber \\
 P_{\phi_1}&=&
\frac{\lambda_1}{\alpha'}
[-\sin^2\theta_1^c\nu_1-\cos^2\theta_1^c
n_1] \ ,
\nonumber \\
 P_{\phi_2}&=&
\frac{\lambda_2}{\alpha'}
[-\sin^2\theta_2^c\nu_1-\cos^2\theta_2^c
n_2] \ .
\nonumber \\
\end{eqnarray}
Inverting these relations we can
express $\omega_1,\omega_2,m_1$ and
$m_2$ as functions  of conserved charges
\begin{eqnarray}
\omega_1=-\frac{\alpha'}{\lambda_1}
[P_{\psi_1}+P_{\phi_1}] \ , \quad
m_1=\frac{\alpha'}{\lambda_1\cos^2\theta_1^c}
[P_{\psi_1}\sin^2\theta_1^c-P_{\phi_1}\cos^2\theta_1^c]
\ ,
\nonumber \\
\omega_2=-\frac{\alpha'}{\lambda_1}
[P_{\psi_2}+P_{\phi_2}] \ , \quad
m_2=\frac{\alpha'}{\lambda_2\cos^2\theta_2^c}
[P_{\psi_2}\sin^2\theta_2^c-P_{\phi_2}\cos^2\theta_2^c]
\ .
\nonumber \\
\end{eqnarray}
Then using (\ref{Tss}) we find the
dependence of energy on the conserved
charges in the form
 \begin{eqnarray}\label{Ed}
E^2&=&P_r^2+P_y^2+
\frac{1}{\lambda_1}(P_{\psi_1}+
P_{\phi_1})^2+
\frac{1}{\lambda_2}(P_{\psi_2}+
P_{\phi_2})^2+ \nonumber \\
&+&\frac{\lambda_1(2\pi m_1)^2}
{(2\pi\alpha')^2}+ \frac{\lambda_2(2\pi
m_2)^2}
{(2\pi\alpha')^2} \ ,  \nonumber \\
\end{eqnarray}
where $m_1=-m_2$, $P_r$ is  radial
momentum \footnote{Note that the motion
of the classical string along radial
direction in I-brane background is free
when we impose conformal gauge.}
 and where
$P_{\phi_2},P_{\psi_2}$ are related to
$P_{\phi_1}$ and $P_{\psi_1}$ as
\begin{eqnarray}\label{angrel}
P_{\phi_2}&=&
\frac{\lambda_2}{\lambda_1}P_{\phi_1}
(\sin^2\theta_2^c-\cos^2\theta_2^c) +
P_{\psi_1}\frac{\lambda_2}{\lambda_1} \left(
\frac{\sin^2\theta_2^c\cos^2\theta_1^c+
\cos^2\theta_2^c\sin^2\theta_1^c}{\cos^2\theta_1}\right)
\nonumber \\
P_{\psi_2}&=&2\frac{\lambda_2}{\lambda_1}
P_{\phi_1}\cos^2\theta_2^c
+\frac{\lambda_2}{\lambda_1}P_{\psi_1}
\frac{\cos^2\theta_2^c}{\cos^2\theta_1^c}
\left(\cos^2\theta_1^c-\sin^2\theta_1^c\right) \ .
\nonumber \\
\end{eqnarray}
 We see that in spite of the fact that
angular momenta are related through the
relations (\ref{angrel}) the dispersion
relation between energy and conserved
charges takes rather simple
form.
 Then it would be really
interesting to identify corresponding
states in dual  I-brane world-volume
theory.
\section{D1-brane in I-brane Background}
\label{fourth} In this section we
generalize discussion presented in
previous section to the case of
D1-brane that moves in I-brane
background. Recall that the dynamics of
probe D1-brane in general background is
governed by the action
\begin{eqnarray}\label{actD1}
S&=&S_{DBI}+S_{WZ} \ , \nonumber \\
S_{DBI}&=&-\tau_1
\int d^2\xi e^{-\Phi}
\sqrt{-\det\bA} \ , \nonumber \\
\bA_{\alpha\beta}&=&\partial_\alpha x^M\partial_\beta x^N
g_{MN}+(2\pi\alpha')\mF_{\alpha\beta}  \ , \nonumber \\
  \mF_{\alpha\beta}&=&
\partial_\alpha A_\beta-\partial_\beta A_\alpha-
(2\pi\alpha')^{-1}b_{MN}\partial_\alpha
x^M\partial_\beta x^N
 \ , \nonumber \\
\end{eqnarray}
where $\tau_1=\frac{1}{2\pi\alpha'}$ is
D1-brane tension,
$\xi^\alpha,\alpha=0,1,
\xi^0\equiv\tau, \xi^1=\sigma$ are
world-volume coordinates and where
$A_\alpha$ is gauge field living on the
world-volume of D1-brane.

If we now perform the variation of (\ref{actD1}) with
respect to $x^M$ we obtain following
equations of motion for  $x^M$
\begin{eqnarray}\label{eqxm}
& &-\tau_1\partial_M[ e^{-\Phi}]
\sqrt{-\det\bA}\nonumber \\
&-&\frac{\tau_1}{2}
e^{-\Phi}(\partial_M g_{KL}\partial_\alpha x^K
\partial_\beta x^L-
\partial_M b_{KL}
\partial_\alpha x^K\partial_\beta x^L)\bAi^{\beta\alpha}
\sqrt{-\det\bA}+\nonumber \\
&+&\tau_1\partial_\alpha
[e^{-\Phi}g_{MN}\partial_\beta x^N\bAi_S^{\beta\alpha}\sqrt{-\det\bA}]-
\nonumber \\
&-&\tau_1
 \partial_\alpha
[e^{-\Phi}b_{MN}\partial_\beta x^N
\bAi_A^{\beta\alpha}\sqrt{-\det\bA}]=0 \ . \nonumber \\
\end{eqnarray}
In the same way the variation of (\ref{actD1})
with respect to  $A_\alpha$ implies following
equation of motion
\begin{equation}\label{eqa}
\partial_\alpha [e^{-\Phi}\bAi^{\beta\alpha}_A\sqrt{-\det\bA}]
=0 \ .
\end{equation}
Now we concentrate on dynamics of
D1-brane in I-brane background and
consider ansatz
\begin{eqnarray}\label{ansD1}
t=\kappa \tau \ , \quad
r=r(\tau)\ , \quad y=y(\tau) \ , \nonumber \\
\psi_1=\omega_1\tau+n_1\sigma \ , \quad \phi_1=
\nu_1\tau+m_1\sigma \ , \nonumber \\
\psi_2=\omega_2\tau+n_2\sigma \ , \quad
\phi_2=\nu_2\tau+m_2\sigma \
\nonumber \\
\end{eqnarray}
and hence  the matrix elements
$\bA_{\alpha\beta}$ take the form
\begin{eqnarray}
\bA_{\tau\tau}&=&-\kappa^2+\dot{r}^2+\dot{y}^2+
\lambda_1\sin^2\theta_1\nu_1^2+\lambda_1\cos^2\theta_1\omega^2_1+
\lambda_2\sin^2\theta_2\nu_2^2+\lambda_2\cos^2\theta_2\omega^2_2
\nonumber \\
\bA_{\sigma\sigma}&=&\lambda_1\sin^2\theta_1
m_1^2+ \lambda_2\cos^2\theta_1 n_1^2+
\lambda_2\sin^2\theta_2
m_2^2+\lambda_2\cos^2\theta_2
n_2^2 \ , \nonumber \\
\bA_{\tau\sigma}&=& \lambda_1
\sin^2\theta_1 \nu_1
m_1+\lambda_1\cos^2\theta_1 \omega_1
n_1+\lambda_2\sin^2\theta_2\nu_2 m_2+
\lambda_2\cos^2\theta_2 \omega_2 n_2-
\nonumber \\
&-&\lambda_1 \cos^2\theta_1(\nu_1
n_1-\omega_1 m_1)- \lambda_2
\cos^2\theta_2(\nu_2 n_2-\omega_2 m_2)+
(2\pi\alpha')F \ ,  \nonumber \\
\bA_{\sigma\tau}&=& \lambda_1
\sin^2\theta_1 \nu_1
m_1+\lambda_1\cos^2\theta_1 \omega_1
n_1+\lambda_2\sin^2\theta_2\nu_2 m_2+
\lambda_2\cos^2\theta_2 \omega_2 n_2+
\nonumber \\
&+&\lambda_1 \cos^2\theta_1(\nu_1
n_1-\omega_1 m_1)+ \lambda_2
\cos^2\theta_2(\nu_2 n_2-\omega_2 m_2)-
(2\pi\alpha')F  \ ,\nonumber \\
\end{eqnarray}
where $F\equiv F_{\tau\sigma}, \quad
\dot{r}=\partial_\tau r,\quad \dot{y}=\partial_\tau y$.
Again, it is easy to see that the time dependence
of $y$ is $y(\tau)=y_0+v_y\tau$.

As the next step we use the fact that
(\ref{eqa}) implies
 an existence of conserved flux $\Pi$
\begin{equation}\label{defPi}
\frac{e^{-\Phi}
(\bA_{\tau\sigma})^A}{\sqrt{-\det\bA}}=\Pi
\ .
\end{equation}
Then, after some algebra, we obtain
\begin{eqnarray}
\sqrt{-\det\bA}&=&
\frac{1}{\sqrt{1+e^{-2Qr}g_s^2\Pi^2}}
[-\dot{r}^2\bA_{\sigma\sigma}+\bM] \ ,
\nonumber \\
\bM&=&-\bA'_{\tau\tau}\bA_{\sigma\sigma}+(\bA_{\tau\sigma})^S
(\bA_{\tau\sigma})^S \ , \quad
\bA'_{\tau\tau}=\bA_{\tau\tau}-\dot{r}^2
\ ,\nonumber \\
\end{eqnarray}
where $\bA'_{\tau\tau}$ and  $\bM$ are
constants.

In order to find the time dependence of
$r$ let us consider the equation of
motion for $x^0=t$ that for the ansatz
(\ref{ansD1}) implies an existence of
the conserved quantity
\begin{equation}\label{defA}
\frac{e^{-\Phi}\bA_{\sigma\sigma}}{\sqrt{-\det
\bA}}=A \ , \quad \mathrm{A}=const \ .
\end{equation}
Then using (\ref{defA}) we easily
determine the differential equation for
$r$
\begin{eqnarray}\label{x0}
\dot{r}=
\sqrt{\frac{\bM}{\bA_{\sigma\sigma}}
-\frac{\bA_{\sigma\sigma}\Pi^2}{A^2}}
\sqrt{1-\frac{\bA_{\sigma\sigma}^2}
{\bM A^2-\bA_{\sigma\sigma}^2\Pi^2}
\frac{e^{2Qr}}{g_s^2}} \ . \nonumber \\
\end{eqnarray}
The differential equation given above
can be  easily integrated  and we
obtain the result
\begin{equation}
\frac{1}{\cosh \left( \tau
Q\sqrt{\frac{\bM
A^2-\bA_{\sigma\sigma}^2\Pi^2}
{\bA_{\sigma\sigma}A^2}}\right)}=
\sqrt{\frac{(\bM
A^2-\bA_{\sigma\sigma}^2\Pi^2)g_s^2}
{\bA_{\sigma\sigma} A^2}}e^{Qr} \ ,
\end{equation}
where we imposed the initial condition
that  at time $\tau_0=0$ D1-brane is
localized in its turning point where
$\dot{r}=0$.

As the next step we solve the equations
of motion for $\phi_{1,2},\psi_{1,2}$.
For example, let us consider the ansatz
(\ref{ansD1}) in the equation of
motion for $\phi_1$
\begin{eqnarray}
& &\partial_\alpha[e^{-\Phi}
g_{\phi_1\phi_1}\partial_\beta
\phi_1\bAi^{\beta\alpha}_S
\sqrt{-\det\bA}]-
\partial_\alpha[e^{-\Phi}
b_{\phi_1\psi_1}\partial_\beta \psi_1\bAi^{\beta\alpha}_A
\sqrt{-\det\bA}]=\nonumber \\
&=&\nu_1
\partial_\tau[g_{\phi_1\phi_1}
\frac{e^{-\Phi}\bA_{\sigma\sigma}}
{\sqrt{-\det\bA}}]
+n_1\partial_\tau[ b_{\phi_1\psi_1}
\frac{e^{-\Phi}
(\bA_{\tau\sigma})^A}{\sqrt{-\det\bA}}]=0
 \ ,  \nonumber \\
\end{eqnarray}
where in the first step we used the
fact  that $g_{\phi_1\phi_1}$ is
constant for constant $\theta_1$, in
the second step  the fact that
$\det\bA$ is a function of $\tau$ only
and in the final step  (\ref{defA}) and
(\ref{defPi}). In the same way we can
show that the equations of motion for
$\psi_1,\phi_2,\psi_2$ are satisfied.

Let us now  analyze the
equations of motion for
$\theta_1,\theta_2$.  For constant
$\theta_1$ its  equation
of motion  takes the form
\begin{eqnarray}\label{theta1eq}
& &
\frac{e^{-\Phi}}{\sqrt{1+e^{2\Phi}\Pi^2}
\sqrt{-\bA_{\tau\tau}\bA_{\sigma\sigma}
+(\bA_{\tau\sigma})^S(\bA_{\tau\sigma})^S}}\times
\nonumber \\
&\times & \left(-\frac{\delta
\bA_{\sigma\sigma}} {\delta
\theta_1}\bA_{\tau\tau}-
\bA_{\sigma\sigma}\frac{\delta
\bA_{\tau\tau}} {\delta
\theta_1}+2(\bA_{\tau\sigma})^S
\frac{\delta
(\bA_{\tau\sigma})^S}{\delta\theta_1}
\right)=0 \ .
\end{eqnarray}
Since $\bA_{\tau\tau}$ contains
$\dot{r}$ it is manifestly time
dependent. Then in order to obey
(\ref{theta1eq}) we have to demand that
\begin{eqnarray}
\frac{\delta \bA_{\sigma\sigma}}
{\delta\theta_1}&=& 2\lambda_1
\sin\theta_1\cos\theta_1
(m_1^2-n_1^2)=0 \ ,
\nonumber \\
\frac{\delta \bA_{\tau\tau}}
{\delta\theta_1}&=& 2\lambda_1
\sin\theta_1\cos\theta_1
(\omega_1^2-\nu_1^2)=0 \ .
\nonumber \\
\end{eqnarray}
We solve these equations by imposing
following relations
\begin{equation}\label{m1n1d1}
m_1=n_1 \ , \quad \omega_1=\nu_1 \ .
\end{equation}
Then it is easy to see that
$\frac{\delta \bA_{\tau\sigma}}{\delta
\theta_1}=\lambda_1
\sin\theta_1\cos\theta_1 (\nu_1
m_1-\omega_1 n_1)$ is   satisfied for
(\ref{m1n1d1}) as well. In the same way we derive
\begin{equation}\label{m2n2d1}
m_2=n_2  \ , \quad \omega_2=\nu_2
\end{equation}
from the equation of motion for $\theta_2$.
Then we obtain
\begin{eqnarray}
& &
\bM=-\kappa^2+v_y^2+\lambda_1(\omega_1^2+n_1^2)+
\lambda_2(\omega_2^2+n_2^2) \ ,
\nonumber \\
& &
\bA_{\sigma\sigma}=\lambda_1n_1^2+\lambda_2
n_2^2
\ , \nonumber \\
& &(\bA_{\sigma\tau})^S= \lambda_1
\omega_1 n_1+\lambda_2 \omega_2 n_2 \ .
\nonumber \\
\end{eqnarray}
As follows from the form of the I-brane
background D1-brane possesses following
conserved currents
\begin{eqnarray}
\mJ^\alpha_y&=&\frac{\delta \mL}{\delta
\partial_\alpha y}=
-\tau_1 e^{-\Phi}g_{yy}
\partial_\beta y \bAi^{\beta\alpha}_S\sqrt{-\det\bA} \ ,
\nonumber \\
\mJ^\alpha_t&=&
\frac{\delta \mL}{\delta \partial_\alpha t}
=-\tau_1
e^{-\Phi}g_{tt}\partial_\beta t\bAi^{\beta\alpha}_S
\sqrt{-\det\bA} \ , \nonumber \\
\mJ^\alpha_{\phi_1}&=&\frac{\delta \mL}{\delta \partial_\alpha\phi_1}=
-\tau_1 e^{-\Phi}[
g_{\phi_1\phi_1}\partial_\beta \phi_1\bAi^{\beta\alpha}_S
+ b_{\phi_1\psi_1}\partial_\beta \psi_1
\bAi^{\beta\alpha}_A]\sqrt{-\det\bA} \ , \nonumber \\
\mJ^\alpha_{\phi_2}&=&
\frac{\delta \mL}{\delta\partial_\alpha \phi_2}=
-\tau_1 e^{-\Phi}[
g_{\phi_2\phi_2}\partial_\beta \phi_2\bAi^{\beta\alpha}_S
+ b_{\phi_2\psi_2}\partial_\beta \psi_2
\bAi^{\beta\alpha}_A]\sqrt{-\det\bA} \ , \nonumber \\
\mJ^\alpha_{\psi_1}&=&\frac{\delta \mL}{\delta \partial_\alpha\psi_1}=
-\tau_1 e^{-\Phi}[
g_{\psi_1\psi_1}\partial_\beta \psi_1\bAi^{\beta\alpha}_S
+ b_{\psi_1\phi_1}\partial_\beta \phi_1
\bAi^{\beta\alpha}_A]\sqrt{-\det\bA} \ , \nonumber \\
\mJ^\alpha_{\psi_2}&=&
\frac{\delta \mL}{\delta\partial_\alpha \psi_2}=
-\tau_1 e^{-\Phi}[
g_{\psi_2\psi_2}\partial_\beta \psi_2\bAi^{\beta\alpha}_S
+ b_{\psi_2\phi_2}\partial_\beta \psi_2
\bAi^{\beta\alpha}_A]\sqrt{-\det\bA} \ .  \nonumber \\
\end{eqnarray}
These currents are
conserved as a
consequence of
the equation of motion:
\begin{equation}
\partial_\alpha \mJ^\alpha_{x}=0 \ , \quad  x=
(t,\phi_1,\phi_2,\psi_1,\psi_2) \ .
\end{equation}
Then   corresponding conserved charges
take the form
\begin{eqnarray}
P_t&=&\int_0^{2\pi}d\sigma \mJ^\tau_t \ ,
\quad P_y=\int_0^{2\pi}d\sigma \mJ^\tau_y \ ,
\nonumber \\
P_{\phi_{1,2}}&=&\int_0^{2\pi}d\sigma
\mJ^\tau_{\phi_{1,2}} \ , \quad
P_{\psi_{1,2}}=\int_0^{2\pi}d\sigma
\mJ^\tau_{\psi_{1,2}} \nonumber \\
\end{eqnarray}
that, for the ansatz (\ref{ansD1}) are equal to
\begin{eqnarray}
P_t&=&-\tau_1 2\pi \kappa A \ ,
\quad
P_y=\tau_1 2\pi v_y A  \ ,
 \nonumber \\
P_{\phi_1}&=& 2\pi\tau_1 \sin^2\theta_1
A
\frac{\lambda_1\lambda_2(n_2\omega_1-\omega_2
n_1)n_2} {\lambda_1 n_1^2+\lambda_2
n_2^2} -2\pi\tau_1
\lambda_1\cos^2\theta_1 n_1 \Pi \ ,
\nonumber \\
P_{\psi_1}&=& 2\pi\tau_1 \cos^2\theta_1
A
\frac{\lambda_1\lambda_2(n_2\omega_1-\omega_2
n_1)n_2} {\lambda_1 n_1^2+\lambda_2
n_2^2} +2\pi\tau_1
\lambda_1\cos^2\theta_1 n_1 \Pi \ ,
\nonumber \\
P_{\phi_2}&=&-2\pi\tau_1 \sin^2\theta_2
A
\frac{\lambda_1\lambda_2(n_2\omega_1-\omega_2
n_1)n_1} {\lambda_1 n_1^2+\lambda_2
n_2^2} -2\pi\tau_1
\lambda_2\cos^2\theta_2 n_2 \Pi \ ,
\nonumber \\
P_{\psi_2}&=& -2\pi\tau_1
\cos^2\theta_2 A
\frac{\lambda_1\lambda_2(n_2\omega_1-\omega_2
n_1)n_1} {\lambda_1 n_1^2+\lambda_2
n_2^2} +2\pi\tau_1
\lambda_1\cos^2\theta_2 n_2 \Pi \ .
\nonumber \\
\end{eqnarray}
The remarkable property of the solution
given above is that  there is not any
relation between energy and angular
momenta. Remember that in case of the
fundamental string an existence of this
relation follows from the Virasoro
constraints. Interestingly we can find
similar relation when we impose the
condition $\dot{r}=0$. In fact, as
follows from (\ref{x0}) this situation
occurs when
\begin{eqnarray}
A^2(-\bA'_{\tau\tau}\bA_{\sigma\sigma}
+(\bA_{\tau\sigma})^S(\bA_{\tau\sigma})^S)
-\bA_{\sigma\sigma}^2\Pi^2=0 \ .
\nonumber \\
\end{eqnarray}
The equation above can be solved with
the ansatz
\begin{equation}\label{gfc}
\bA'_{\tau\tau}=\bA_{\tau\tau}=-\bA_{\sigma\sigma}
\ , \quad  (\bA_{\tau\sigma})^S=0 \
, \quad \Pi^2=A^2 \ ,
\end{equation}
where we used the fact that for
$\dot{r}^2=0, \
\bA'_{\tau\tau}=\bA_{\tau\tau}$. Using
these prescriptions we derive
constraints that have  similar form as
 Virasoro constraints for fundamental
string.

Further, using (\ref{gfc}) it is easy
to see that  the equation
of motion for $\theta_1$ is solved with
the ansatz (\ref{m1n1d1}) and the equation
of motion for $\theta_2$ with the ansatz
(\ref{m2n2d1}).
Then the condition
$(\bA_{\tau\sigma})^S=0$ implies
\begin{equation}\label{omega12d1}
\lambda_1 \omega_1n_1+
\lambda_2 n_2\omega_2=0 \
\end{equation}
that can be also solved as
\begin{equation}\label{omega21d}
\omega_2=\omega_1 \frac{\lambda_1}{\lambda_2} \ , \quad
n_1=-n_2 \
\end{equation}
that has again the same form as the
relations derived in previous section.
Then, for (\ref{omega12d1}) and
for  $A=\Pi$
 we obtain
\begin{eqnarray}
P_t&=&
-\tau_1 2\pi \Pi \kappa
 \ , \quad P_y=\tau_1 2\pi \Pi v_y \ ,
\nonumber \\
P_{\phi_1}&=&
2\pi\tau_1 \Pi \lambda_1
[\sin^2\theta_1 \omega_1-\cos^2\theta_1
m_1] \ ,
\nonumber \\
P_{\psi_1}
&=&2\pi\tau_1 \Pi \lambda_1
[\cos^2\theta_1 \omega_1+\cos^2\theta_1  m_1] \ ,
\nonumber \\
P_{\phi_2}&=& 2\pi\tau_1 \Pi \lambda_2
[\sin^2\theta_2 \omega_2-\cos^2\theta_2
m_2] \ ,
\nonumber \\
P_{\psi_2}&=& 2\pi\tau_1 \Pi \lambda_2
[\sin^2\theta_2 \omega_2+
\cos^2\theta_2  m_2] \ ,
\nonumber \\
\end{eqnarray}
where again $\omega_2,m_2$ are related
to $\omega_1,m_1$ through the relations
(\ref{omega21d}) that allow to express
$P_{\phi_2},P_{\psi_2}$ as functions of
$P_{\phi_1},P_{\psi_1}$ exactly in the
same way as in previous section.
 Further, the condition
\begin{equation}
\bA_{\tau\tau}+\bA_{\sigma\sigma}=0
\end{equation}
implies
\begin{equation}
\kappa^2=v_y^2+\lambda_1(\omega_1^2+n_1^2)+
\lambda_2(\omega_2^2+n_2^2)
\end{equation}
and hence
\begin{eqnarray}
E^2&=&P_y^2+\frac{1}{\lambda_1}
(P_{\phi_1}+P_{\psi_1})^2+
\lambda_1(2\pi\tau_1 \Pi n_1)^2+
\nonumber \\
&+&\frac{1}{\lambda_2}
(P_{\phi_2}+P_{\psi_2})^2+
\lambda_2(2\pi\tau_1 \Pi n_2)^2 \ .
\nonumber \\
\end{eqnarray}
The relation given above has similar
form as the relation that was derived
in previous section for special case
$P_r=0$. Since $\Pi$ determines the
number of fundamental strings that
propagate on the world-volume of
D1-brane we can interpret the solution
given above as a bound state of $\Pi$
fundamental string and one D1-brane
that is localized in radial direction.
It is a great challenge to find
corresponding  interpretation of this
configuration in dual holographic
theory.

 {\bf Acknowledgement}

This work
 was supported  by the Czech Ministry of
Education under Contract No. MSM
0021622409.


\end{document}